\documentclass[a4paper,11pt]{article}
\usepackage{graphicx,url,amssymb,amsmath,rotating,color,units,wasysym,epsfig,multirow,epstopdf,subcaption,units}
\usepackage{jcappub}
\usepackage{soul,xcolor}
\usepackage{siunitx}
\usepackage{url}
\usepackage{widetext}
\graphicspath{{figures/}}
\begin{document}
\title{Bayesian WIMP detection with the Cherenkov Telescope Array}

\author[a]{Abhi Mangipudi,}
\author[a]{Eric Thrane}
\author[a]{and Csaba Balazs}
\affiliation[a]{School of Physics and Astronomy, Monash University, Melbourne VIC 3800, Australia}
\emailAdd{abhi.mangipudi@monash.edu}
\emailAdd{eric.thrane@monash.edu}
\emailAdd{csaba.balazs@monash.edu}

%\author{Abhi Mangipudi}
%\email{abhi.mangipudi@monash.edu}
%\author{Eric Thrane}
%\email{eric.thrane@monash.edu}
%\author{Csaba Balazs}
%\email{csaba.balazs@monash.edu}
%\affiliation{School of Physics and Astronomy, Monash University, Melbourne VIC 3800, Australia}

\abstract{
Over the past decades Bayesian methods have become increasingly popular in astronomy and physics as stochastic samplers have enabled efficient investigation of high-dimensional likelihood surfaces.
In this work we develop a hierarchical Bayesian inference framework to detect the presence of dark matter annihilation events in data from the Cherenkov Telescope Array (CTA). 
Cosmic rays are weighted based on their measured sky position $\hat\Omega_m$ and energy $E_m$ in order to derive a posterior distribution for the dark matter's velocity averaged cross section $\langle\sigma v\rangle$.
The dark matter signal model and the astrophysical background model are cast as prior distributions for $(\hat\Omega_m, E_m)$.
The shape of these prior distributions can be fixed based on first-principle models; or one may adopt flexible priors to include theoretical uncertainty, for example, in the dark matter annihilation spectrum or the astrophysical distribution of sky location.
We demonstrate the utility of this formalism using simulated data with a Galactic Centre signal from scalar singlet dark-matter model.
The sensitivity according to our method is comparable to previous estimates of the CTA sensitivity.}

\maketitle

\section{Introduction}\label{intro}
The Cherenkov Telescope Array (CTA) is well equipped to detect dark-matter annihilation or decay events, probing parameters of the weakly interacting massive particle (WIMP) paradigm \cite{cta}. There is a reasonable chance that dark matter belongs to the WIMP class \cite{bertone}, and is a thermal relic of the early Universe \cite{berg, GAMBIT:2021rlp, GAMBIT:2018gjo, GAMBIT:2018eea, Athron:2018ipf, GAMBIT:2017snp, GAMBIT:2017zdo, GAMBIT:2017gge}.
The cold % $(v \ll c)$ strictly, v is undefined here
dark matter paradigm has proven to be successful in reproducing most observations down to galactic scales with some open questions remaining at smaller scales \cite{buckley}. With collider and direct detection searches yet to yield universally accepted, statistically significant signals~\footnote{While the purported detection of dark matter particles by DAMA/LIBRA is statistically significant \cite{Bernabei2008}, this claim is controversial.}, recent years have seen increasing interest in indirect detection methods. With plans of observing the Galactic Centre and some dwarf spheroidal satellite galaxies, CTA will be sensitive to the WIMP thermal cross-section for a dark matter mass in the range $\sim$ 200 GeV to 20 TeV \cite{cta}.

Indirect detection experiments like CTA can determine $\langle \sigma v\rangle$---the velocity-weighted average annihilation cross section---by measuring the flux of gamma rays originating from dark matter annihilation.
For dark matter that is a thermal relic of the early Universe with mass of the electroweak scale, the expected value of $\langle\sigma v \rangle$ is approximately \cite{ros},
\begin{equation}\label{eq:thermalcross}
\langle \sigma v\rangle \approx \unit[3\times 10^{-26}]{cm^3s^{-1}} .
\end{equation}
Here, $v$ is the relative velocity of the colliding dark matter particles, $\sigma$ is the total annihilation cross section and the angled brackets denote an ensemble average.
The velocity-weighted annihilation cross-section is related to the differential gamma-ray flux from dark matter annihilation:
\begin{align}\label{eq:flux}
    \frac{d\Phi}{dE} = 
    \frac{1}{4 \pi}
    J
%    \bigg(
    \frac{\langle\sigma v\rangle}{2S_{\chi} m_{\chi}^2} \sum_f B_f \frac{dN_f}{dE} .
%    \bigg) .
\end{align}
Here, $d\Phi/ dE$ is the gamma-ray flux from dark-matter annihilation.
The variable $J$ is the so-called $J$-factor,
\begin{align} \label{eq:jfactor}
    J = \int d\hat{n}\int_\text{LOS}  \, ds \, \rho_{\chi}^2(r(s,\hat{n})),
\end{align}
which depends on the dark-matter density $\rho_\chi(r(s,\hat{n}))$ along the line of sight (LOS). The integral is over the unit vector $\hat{n}$ and $s$ is the distance along the LOS.
When observing the Galactic Centre the dark matter density is typically assumed to follow the Einasto profile \cite{einasto}:
\begin{align} \label{eq:einasto}
\rho_{\chi}(r) = \rho_0 \exp\bigg\{-\frac{2}{\alpha} \bigg[ \bigg(\frac{r}{r_s}\bigg)^{\alpha} - 1\bigg]\bigg\}.
\end{align}
Here $r_s$ and $\rho_0$ are the radius and density, respectively, at which the logarithmic slope of the density is $-2$, and $\alpha$ is a parameter describing the degree of curvature of the profile \cite{viana}. We take $r_s = \unit[20]{kpc}$, $\alpha = 0.17$ \cite{einasto}. The parameter $\rho_0$ is fixed such that the local dark matter density is $\rho_{\chi} (r_{\odot}) = \unit[0.39]{GeV \, cm^{-3}}$, where $r_{\odot} = \unit[8.5]{kpc}$ \cite{Catena}.

Returning to Eq.~\ref{eq:flux}, the mass of the dark matter particle is $m_\chi$, and $S_{\chi} = 1 (2)$ if this particle is (not) its own anti-particle.  The sum is over annihilation states $f$, $B_f$ is the annihilation fraction while $dN_f/dE$ is the photon energy spectrum. In factorising the $J$-factor from the terms in parentheses we assume that the annihilation rate is uncorrelated with $v(r)$ \cite{CTAprecon}.

It is instructive to pause and discuss the components of Eq.~\ref{eq:flux}.
The left-hand side, the differential gamma-ray flux, is what experiments like CTA measure by looking for excess photons beyond what is expected from the astrophysical background.
If we assume a specific model for dark-matter annihilation, most of the terms on the right-hand side can be calculated from first principles.
The one right-hand-side term that we cannot uniquely calculate from first principles is $\langle \sigma v \rangle$.
Thus, by measuring the differential flux, and assuming a specific dark-matter model, we obtain an estimate for $\langle \sigma v \rangle$.
The main mission of indirect detection experiments is to measure $m_\chi$ and $\langle \sigma v \rangle$.

%This annihilation rate includes contributions from various partial waves of the scattering amplitude, each with a varying dependence on the relative velocity $v$ of the colliding particles,
%\begin{equation}
%\langle\sigma v\rangle=\underbrace{a}_{\langle\sigma v\rangle_{s}}+\underbrace{b v^{2}}_{\langle\sigma v\rangle_{p}}+\ldots
%\end{equation}
%The first term on the RHS, $\langle\sigma v\rangle_s$, represents the velocity independent $s$-wave contribution and the second term, $\langle\sigma v\rangle_p$, which scales as $v^2$, is the $p$-wave contribution. We have omitted terms of higher powers of $v^2$, since these contributions are negligible for cold dark matter models. If the $s$-wave contribution dominates ($b \gg a$) then the value of $\langle \sigma v\rangle$ today can be similar to the value at freeze-out because there is no $v$ dependence. For $p$-wave annihilation ($b \gg a$), and so the value of $\langle \sigma v\rangle$ today  is much larger than at freeze out.
%Since $\langle \sigma v\rangle$ at freeze out is fixed by the dark matter abundance (see Eq. 1), such models would be out of reach for indirect detection methods \cite{das1}. Furthermore, Eq (1) places an upper bound of $m_\chi\lesssim \unit[100]{TeV}$, as higher masses than this would violate unitarity. We can also place a lower bound of around $\unit[200]{GeV}$, as below this mass, previous indirect searches would likely have made a detection.

A number of previous publications have explored how CTA will be able to measure $\langle \sigma v\rangle$; see, for example, \cite{CTAprecon, viana, Lefranc, Hutten, Pierre, Silverwood, Aguirre-Santaella}.  Ref. \cite{CTAprecon} uses the on/off approach to demarcate two separate regions of interest---an `on' region in which the signal is expected to dominate over background, and an `off' region in which the background dominates over signal.  Ref. \cite{viana} utilises a Poisson joint-likelihood method, where the comparisons between dark matter and background fluxes are performed in different energy and spatial bins.  Ref. \cite{Lefranc} uses a likelihood ratio test statistic, which uses the log likelihood ratio comparing the signal hypothesis to the ``total'' hypothesis to quantify the statistical significance of the signal, which can in turn be used to estimate $\langle \sigma v\rangle$.

In this work we develop a Bayesian formalism for the measurement of dark matter annihilation in CTA data.
In doing so, we aim to leverage Bayesian tools used widely in astronomy such as stochastic samplers, which can be used to efficiently explore high-dimensional likelihood surfaces.
A Bayesian formulation is also useful pedagogically, for example, to explicitly formulate assumptions as priors.
Our signal model (from dark matter annihilation) and our background model (of gamma rays from astrophysical processes) are framed in terms of prior distributions for sky location $\Omega$ and photon energy $E$.
We construct a posterior distribution for the velocity-averaged cross section $\langle \sigma v \rangle$.
A dark matter signal is detected when this posterior excludes $\langle \sigma v \rangle = 0 $ at high credibility, e.g., 99.99994\% for a ``five-sigma'' detection.
Systematic uncertainty, for example, in the shape of our signal and background distributions, can be modeled with hyper-parameters; see, e.g.,~\cite{intro}.
Marginalizing over these hyper-parameters, one obtains a new posterior distribution for $\langle \sigma v \rangle$, which is (appropriately) broadened by systematic uncertainty.

The remainder of this paper is organised as follows.
In Section~\ref{formalism}, we present the statistical formalism underpinning our analysis.
We describe the prior distributions that characterise our signal and background models; we describe the likelihood function characterising CTA measurements of cosmic rays.
In Section~\ref{demo}, we demonstrate the formalism using simulated data.
We provide concluding thoughts in Section~\ref{discussion}.

\section{Formalism}\label{formalism}
\subsection{Overview and basics}
In this section we describe a Bayesian formalism to measure the properties of dark-matter annihilation in CTA data.
We consider a dataset consisting of $N$ gamma-ray events, each indexed by an event number $1 \leq i \leq N$.  The dataset $\vec{d}$, which describes these $N$ events, consists of elements $d_i$.  Each datum $d_i$ consists of the measured sky location $\hat\Omega_m$ and the measured energy $E_m$ of each cosmic ray~\footnote{The ``raw'' data produced from CTA is more complicated than we describe here, consisting of CCD images and arrival times. Our data is the result of pre-processing to convert the raw data into estimates of the sky location and energy of each event.}:
\begin{align}
    d^i = \{\hat\Omega_m^i E_m^i\} .
\end{align}
Here, the hat signals that $\hat\Omega_m$ is a unit vector.

\subsection{Likelihood}
\begin{figure*}[htpb!]
     \centering
     \begin{subfigure}[b]{0.49\textwidth} 
         \centering
         \includegraphics[width=\textwidth]{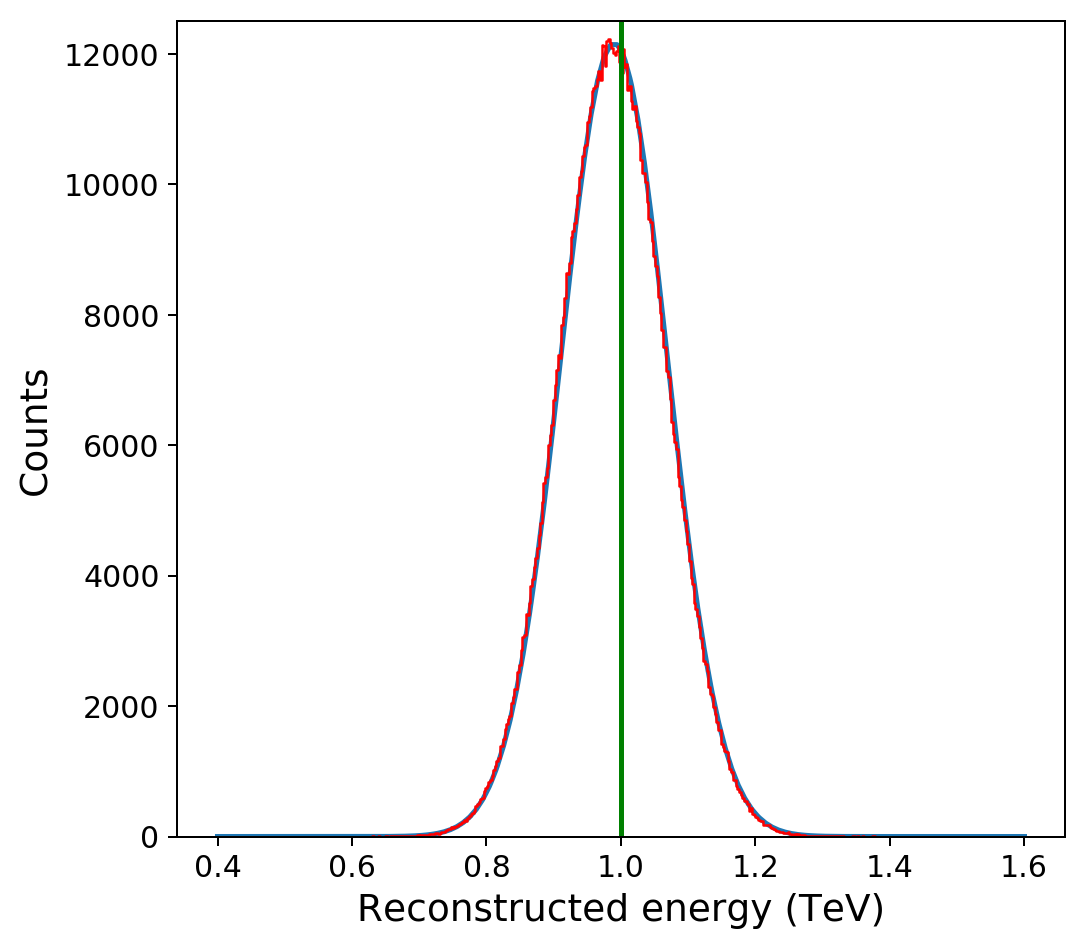}
         \caption{}
     \end{subfigure}
     \hfill
     \begin{subfigure}[b]{0.49\textwidth}
         \centering
         \includegraphics[width=\textwidth]{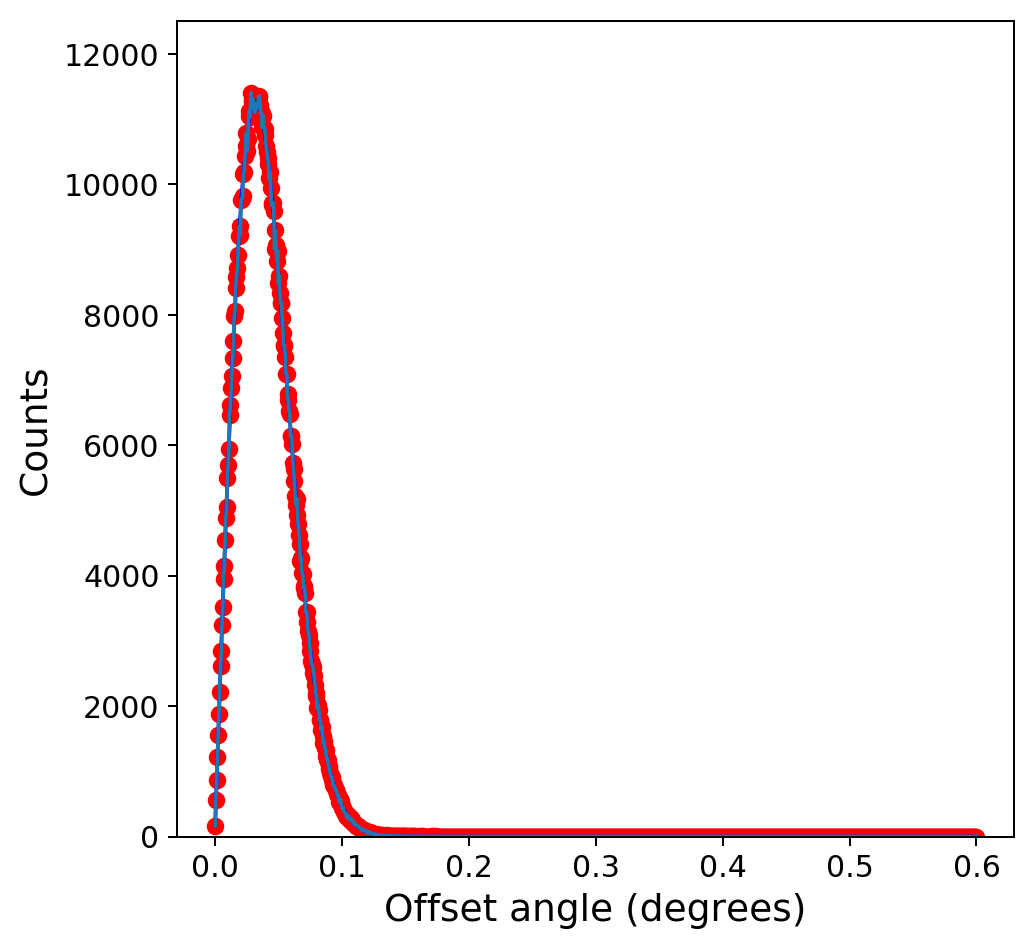}
         \caption{}
     \end{subfigure}
        \caption{Likelihood functions.
        \textit{Left}: example of the likelihood for measured photon energy given true photon energy ${\cal L}(E_m^i | E^i)$. In this case, the true energy (indicated by the vertical green line) is $E=\unit[1]{TeV}$.
        \textit{Right}: example of the point spread function ${\cal L}(\hat\Omega_m^i | \hat\Omega^i, E_m^i)$ for a given true energy $ E = \unit[1]{TeV}$.
        Both examples are produced using \texttt{ctools} functions \texttt{test\_sim\_edisp} \cite{sim_edisp} for energy and \texttt{test\_sim\_psf} \cite{sim_psf} for sky location.}
        \label{fig:psf}
\end{figure*}

Each event $i$ is characterized by a likelihood function ${\cal L}(d^i | \hat\Omega^i, E^i)$.
This likelihood is a point spread function: it relates the \textit{measured} sky location and energy (denoted with a subscript $m$) to the \textit{true} sky location and energy (written without subscripts).
It can be factored into separate components for energy and sky location
\begin{align}\label{eq:likelihood}
    {\cal L}(d^i | \hat\Omega^i, E^i) = 
%    {\cal L}_E(...) \, 
%    {\cal L}_\Omega(...) \nonumber\\
    {\cal L}(E_m^i | E^i) \, 
    {\cal L}(\hat\Omega_m^i | \hat\Omega^i, E_m^i) .
\end{align}
The sky location likelihood is conditional on the measured energy since high-energy cosmic rays are better localised than low-energy ones. We use the publicly available \texttt{prod3b-v1} \footnote{Specifically, we use the \texttt{South\_z20\_50h} IRF file.} instrument response function (IRF) library, which is optimised for the detection of a point-like source at a $20^{\circ}$ zenith angle.  Note that the Galactic Centre is mostly visible from the southern site \cite{CTAprecon}. Our likelihood function uses the \texttt{ctools} \cite{ctools} functions \texttt{test\_sim\_edisp} \cite{sim_edisp} for energy and \texttt{test\_sim\_psf} \cite{sim_psf} for sky location. These likelihood functions are shown in Fig.~\ref{fig:psf}a and~\ref{fig:psf}b respectively.

\subsection{Signal model}
Gamma rays from dark matter annihilation are distinguishable from the astrophysical background based on their tendency to cluster near the gravitational potential well in the Galactic Centre.
The energy spectrum from dark matter annihilation is also expected to differ from the astrophysical spectrum, though, the precise shape depends on the details of the unknown dark matter annihilation physics.
Our signal model therefore consists of two parts: a description of the angular distribution and the energy spectrum.

\begin{figure*}[htpb!]
     \centering
     \begin{subfigure}[b]{0.49\textwidth}
         \centering
         \includegraphics[width=\textwidth]{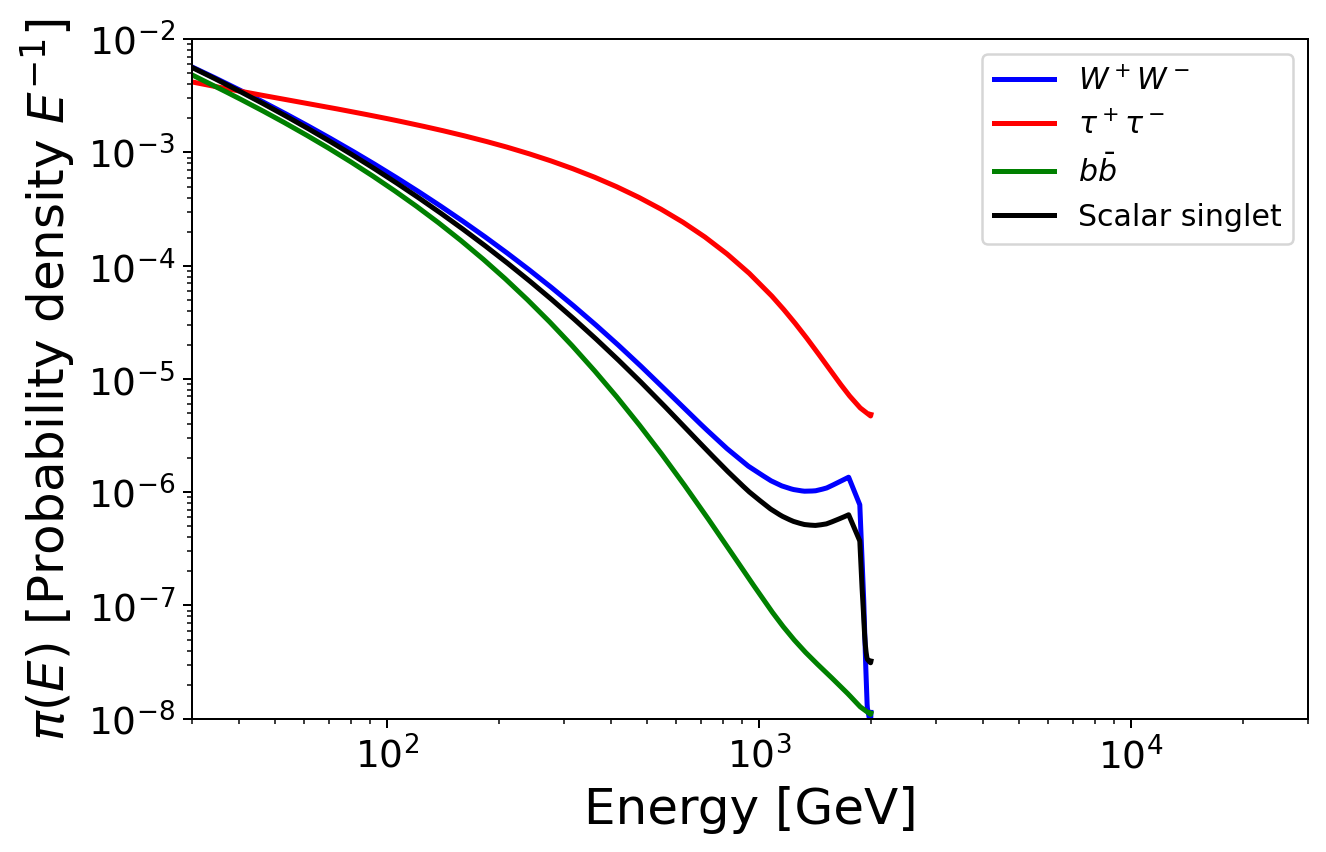}
         \caption{}
     \end{subfigure}
     \hfill
     \begin{subfigure}[b]{0.49\textwidth}
         \centering
         \includegraphics[width=\textwidth]{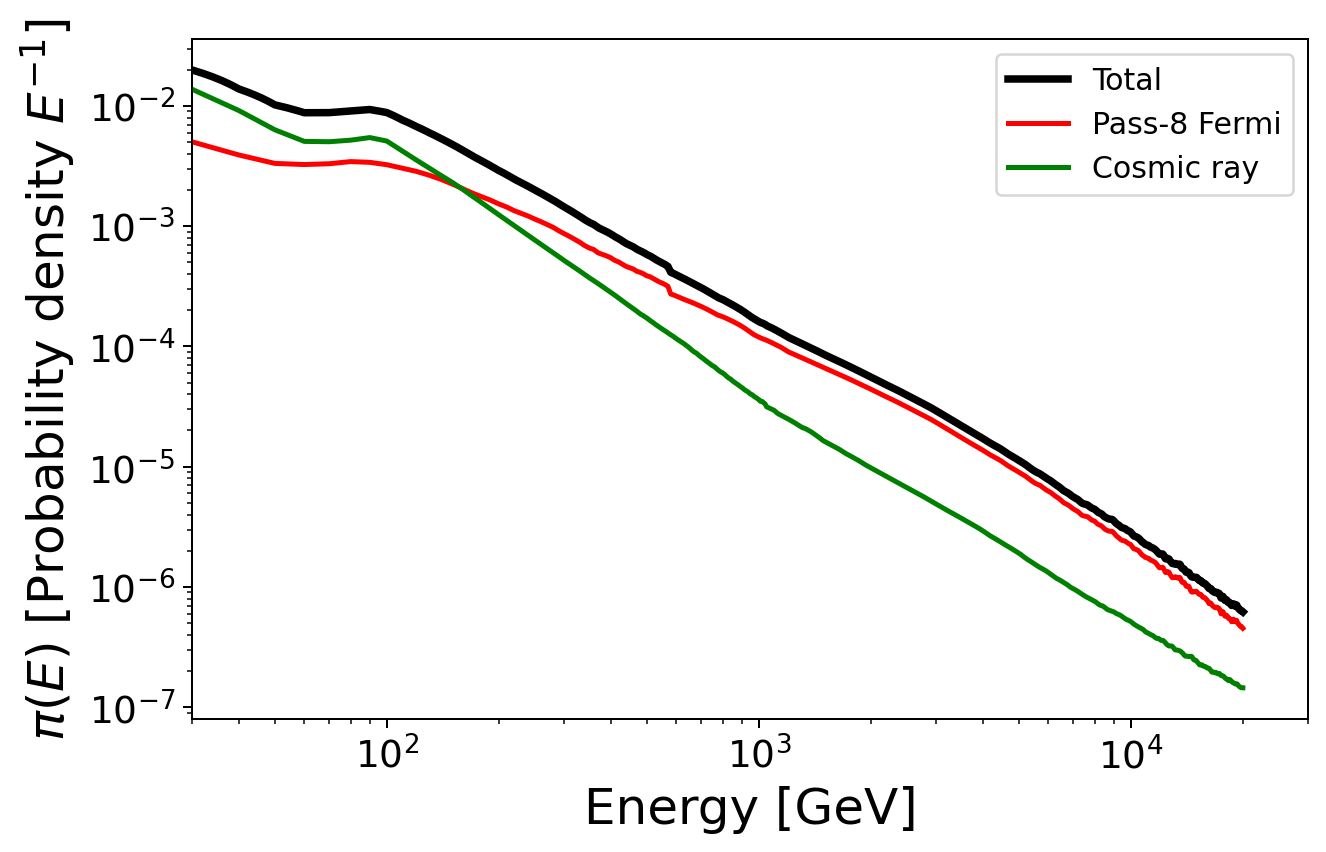}
         \caption{}
     \end{subfigure}
        \caption{
        Prior distributions for energy in our Galaxy. \textit{Left}: example prior distributions for the signal hypothesis $\pi(E^i | {\cal S})$ calculated using the Poor Particle Physicist Cookbook (PPPC) \cite{PPPC}. In this case, the dark matter mass is $m_{\chi} = \unit[2]{TeV}$. We assume a scalar singlet model (black) for which we show sample detection and non-detection plots in Section~\ref{demo}. We also assume a simplified model in which the dark matter particle annihilates to specific standard model final states: $W^+W^-$ (blue), $\tau^+\tau^-$ (red) and $b\bar{b}$ (green).
        \textit{Right}: the prior distribution (black) for the background hypothesis $\pi(E^i | {\cal B})$ using \texttt{ctools}. The black curve includes contributions from misidentified charged cosmic rays (green) and interstellar emission (red) (for which we choose the \textsc{Pass-8 Fermi} model).
        }
        \label{fig:spectra}
\end{figure*}

The signal prior for the energy of event $i$ is denoted
\begin{align}\label{eq:pi(E|S)}
    \pi(E^i | \mathcal{S}) .
\end{align}
The $\mathcal{S}$ signifies that this prior is for the signal hypothesis (that the cosmic ray was created from dark-matter annihilation), which we contrast below with the background hypothesis (that the cosmic ray was created through astrophysical acceleration) denoted $\mathcal{B}$.
For illustrative purposes, we will assume a specific dark matter scenario: the scalar singlet model.
The energy spectrum for this model $\pi(E | \mathcal{S})$ is shown as the black curve in Fig.~\ref{fig:spectra}a.
Comparing the signal distribution to the background distribution in Fig.~\ref{fig:spectra}b (discussed in greater detail below), we see that the signal appears as an excess (above the astrophysical background) around the assumed dark matter mass, in this case, about $\unit[2]{TeV}$.

The scalar singlet model is one of the simplest WIMP models; the model adds one massive real scalar field $S$ to the standard model \cite{holz,burgess,McDonald}. The only renormalizable interaction terms between the singlet and the standard model permitted in the Lagrangian are of the form $S^2H^2$. This leads to a range of possible phenomenological consequences including: thermal production in the early Universe and present-day annihilation signals \cite{Yaguna,Profumo,Arina}, direct detection and $h \rightarrow SS$ decays \cite{Mambrini}. The renormalizable terms involving $S$, permitted by the $Z_2$, gauge, and Lorentz symmetries are,
\begin{align}\label{eq:scalarsinglet}
    \mathcal{L} = \frac{1}{2}\mu_S^2 S^2  + \frac{1}{2}\lambda_{hS} S^2 |H|^2 + \frac{1}{4}\lambda_SS^4 + \frac{1}{2}\partial_{\mu}S\partial^{\mu}S.
\end{align}
The pre-factors are the bare $S$ mass, the Higgs-portal coupling, the $S$ quartic self-coupling, and the last term is the $S$ kinetic term. If the singlet does not obtain a vacuum expectation value, the model only has three free parameters: $\mu^2_S$, $\lambda_{Sh}$ and $\lambda_S$.
In the following section we present an example detection and non-detection plot (Fig.~\ref{fig:scalarsinglet}) for the following choice of parameters: $\lambda_{Sh} = 1$, $\lambda_S = 1$, and $m_{S} = \unit[2]{TeV}$. These parameters are chosen based on constraints upon the scalar singlet model in \cite{Athron:2018ipf}. The branching fractions and standard model final states that go into producing the energy spectrum are:
\begin{align}
    B_f(W^+ W^-) = 0.50, \\
    B_f(Z^0 Z^0) = 0.25, \\
    B_f(hh) = 0.25,
\end{align}
which have been calculated using \texttt{micrOMEGAs} \cite{micromegas}, a code for the calculation of various dark-matter properties. 

We have also assumed simplified models in which the dark matter particle primarily annihilates to specific standard model final states (Fig.~\ref{fig:spectra}a): $W^+W^-$ (blue), $\tau^+\tau^-$ (red) and $b\bar{b}$ (green). 
While scalar singlet dark matter particles mostly annihilate into a $W^+W^-$ final state, they have considerable annihilation fraction to final states that are rarely considered in simplified models.  This will diminish our ability to measure $\langle \sigma v \rangle$ since some dark matter particles may annihilate in such a way that does not produce detectable cosmic rays.  
Varying the mass of the dark matter particle shifts the signal prior $\pi(E | \mathcal{S})$.

\begin{figure*}[htpb!]
     \centering
     \begin{subfigure}[b]{0.475\textwidth}
         \centering
         \includegraphics[width=\textwidth]{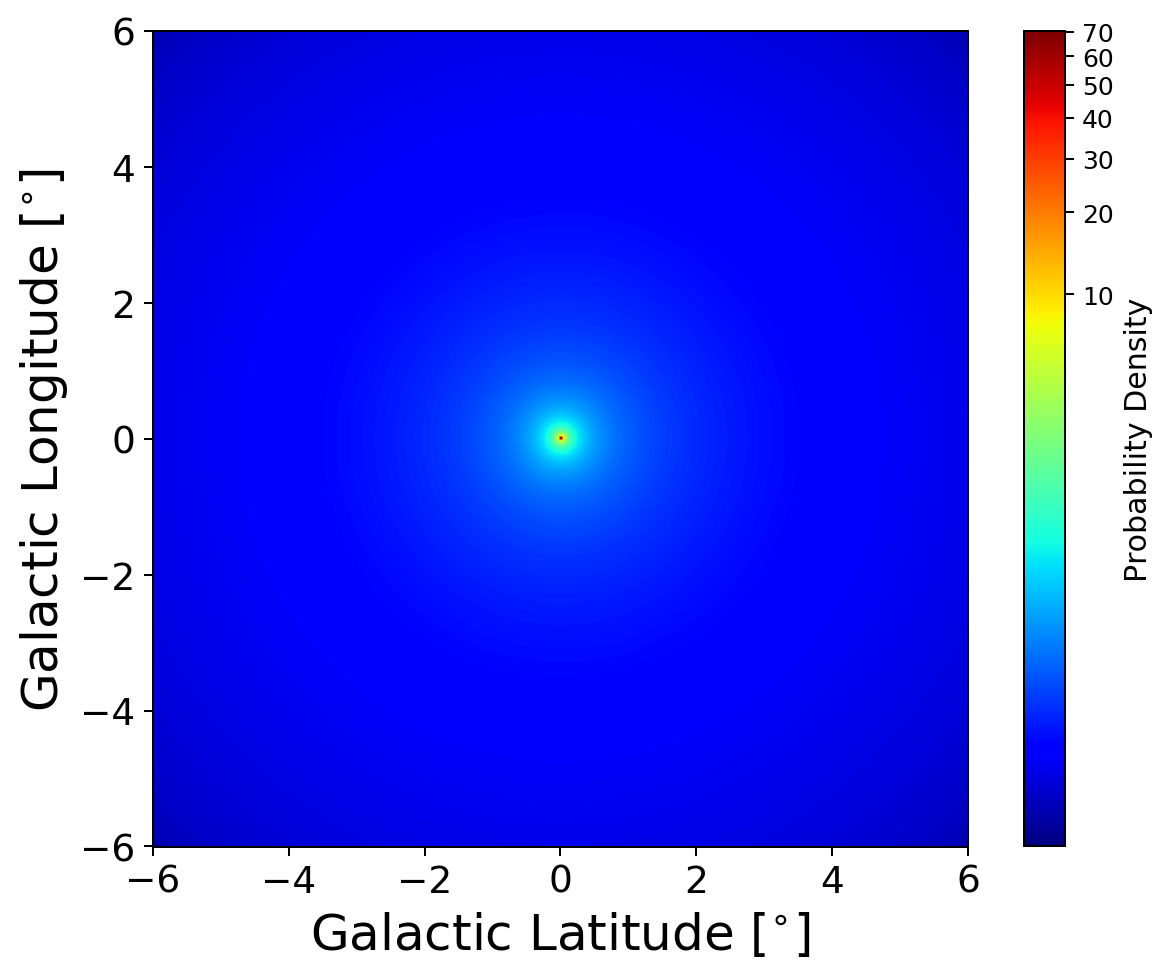}
         \caption{}
     \end{subfigure}
     \hfill
     \begin{subfigure}[b]{0.49\textwidth}
         \centering
         \includegraphics[width=\textwidth]{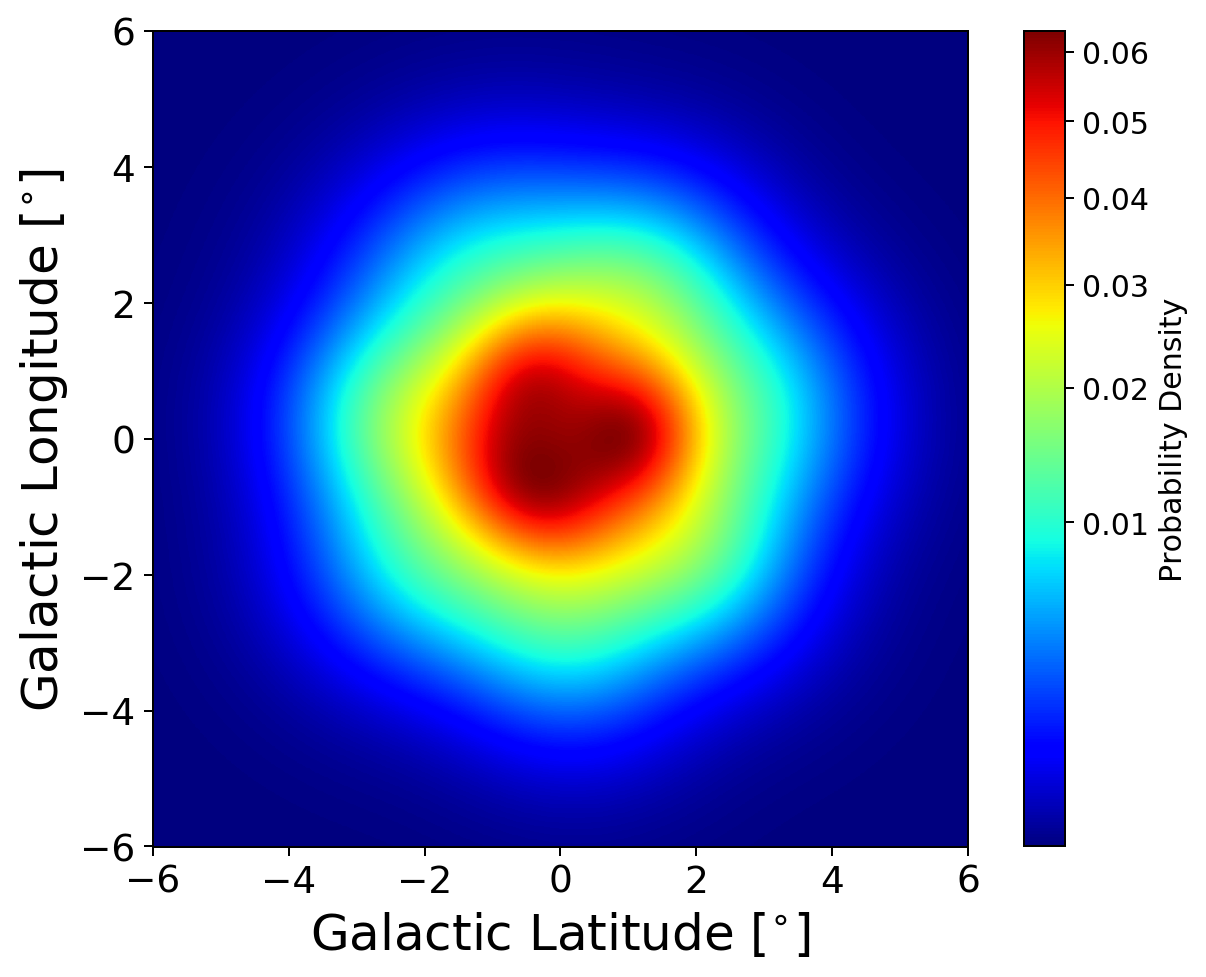}
         \caption{}
     \end{subfigure}
        \caption{Prior distributions for sky location around the Galactic Centre. \textit{Left}: sky location prior for the signal hypothesis $\pi(\Omega^i| {\cal S})$, the Einasto profile plotted in Galactic latitude and longitude $\rho_{\chi}(l,b)$. \textit{Right}: sky location prior for the background hypothesis $\pi(\Omega^i | {\cal B})$. A Gaussian mixture model with ten Gaussians, fitted to \unit[525]{hours} of simulated data of the astrophysical background obtained using \texttt{ctools}.}
        \label{fig:skylocation}
\end{figure*}

The signal prior for the sky location of event $i$ is denoted
\begin{align}
    \pi(\hat\Omega^i | \mathcal{S}) .
\end{align}
We take our signal prior to be the Einasto profile, which is defined above in Eq.~\ref{eq:einasto}.
In Fig.~\ref{fig:skylocation}a, we show the signal prior plotted as a function of Galactic longitude and latitude.
Comparing the signal distribution to the background distribution in Fig.~\ref{fig:skylocation}b (discussed in greater detail below), we see that the signal appears as an excess (above the astrophysical background) near the Galactic Centre.
Some of the most common dark matter density profiles are the Einasto, the NFW (named for the authors, Navarro, Frenk, White) \cite{einasto}, and the Burkert \cite{Burkert}. The Einasto and NFW profiles peak sharply while the Burkert profile levels off \cite{viana}. 
The precise shape of the dark matter distribution near the Galactic Centre is the subject of active research (see, e.g., \cite{Iocco:2016itg}).
Systematic uncertainty in the shape of the dark matter profile leads to uncertainty in the $J$-factors, which propagates to systematic uncertainty in $\langle \sigma v \rangle$ \cite{Benito:2016kyp}. 
This is particularly true for Galactic Center searches \cite{Heros}. The Einasto profile is most consistent with the observed distribution of dark matter \cite{CTAprecon}. However, the dark matter density is relatively poorly constrained and there remain discrepancies between observation and prediction of the dark matter distribution in dwarf galaxies \cite{benito}.
Therefore, while we adopt the Einasto profile here for illustrative purposes, we recommend marginalizing over uncertainty in the dark matter profile.

\subsection{Background model}
Gamma rays are produced in the Galaxy from the decay of relativistic particles accelerated through shocks. When crossing a shock front, these high-energy particles gain energy and scatter off approaching scattering centers. The energy spectrum from shock acceleration follows a power law $dN/dE \propto E^{-n_{\gamma}}$ where $n_{\gamma} \gtrsim 2$ \cite{Blandford, CTAprecon}.
The astrophysical background prior for the energy of event $i$ is denoted
\begin{align}
    \pi(E^i|\mathcal{B}) .
\end{align}
For the Galactic Centre target this distribution is shown in Fig.~\ref{fig:spectra}b. The shape of the spectra appropriately follows the power law described above.
Recent observations of the Galactic Centre indicate an excess of cosmic rays, with dark matter annihilation being a potential explanation \cite{anjos}. The origin of cosmic rays with energies up to $10^{15} \si{eV}$ are commonly attributed to supernova remnants \cite{reynolds}.
However, some other sources have been proposed such as Sagittarius A* and other TeV-bright objects near the galactic centre \cite{fujita, HESS}, and unresolved millisecond pulsars \cite{guepin} all of which are able to accelerate cosmic rays to very high energies.
Additionally, two large gamma ray structures known as ``Fermi Bubbles'' extend above and below the Galactic Centre \cite{cheng}, though, these bubbles (likely sourced by Sagittarius A*) contribute only a small fraction ($\ll 1\%$) of the astrophysical gamma-ray flux.

The background prior for sky location of event $i$ is denoted
\begin{align}
    \pi(\hat\Omega^i | \mathcal{B}) .
\end{align}
In Fig.~\ref{fig:skylocation}b we show the background prior plotted as a function of Galactic longitude and latitude. The shape of the distribution is determined by the interstellar emission which extends along the Galactic plane. Gamma rays along the Galactic plane are predominantly produced through two processes: cosmic ray interactions with the interstellar medium gas (primarily through the neutral pion channel), and cosmic rays up-scattering off the interstellar radiation field and/or the cosmic microwave background photons to gamma ray energies (inverse Compton scattering) \cite{CTAprecon}.

The astrophysical background in our analysis consists of various components.  Following \cite{CTAprecon}, in the background we include a contribution from charged cosmic rays and interstellar emission.  Due to the relatively low expected flux, we neglect the Fermi Bubbles \footnote{Data available at \url{https://github.com/cta-observatory/cta-gps-simulation-paper/tree/master/skymodel/iem}} and point sources.  The contribution from charged cosmic rays is the most dominant background component, which is mostly due to misidentified electrons. Ref. \cite{CTAprecon} uses three different models for the interstellar emission: the \textsc{Gamma}, \textsc{Base+Galactic Ridge}, and \textsc{Pass-8 Fermi} models.  
%
% The Gamma model introduces a radial dependence on cosmic ray diffusion, with diffusion being more efficient closer to the galactic centre. This leads to harder and brighter gamma ray emission in the galactic centre, explaining simultaneously the bright Galactic Ridge in the very centre and the large scale diffuse emission measured by Fermi-LAT. The Base model instead assumes a constant CR diffusion coefficient across the Galaxy. The Base model is also more heavily based on theoretical expectations than other models. The large scale diffuse emission measured by Fermi-LAT and the emission from the Galactic Ridge (H.E.S.S.) must then have a different origin, with the latter theorised to originate from an unknown source related to transient emission from the galactic centre. The interstellar emission model generating Fermi was derived based on an analysis of 8 years of Fermi-LAT data. Special care was taken for the model to describe the high-energy $(\geq \unit[50]{GeV})$ spectrum. It is tuned to the LAT data over the entire sky \cite{CTAprecon}.
%
For a detailed description of these models we refer the reader to Ref. \cite{CTAprecon}.
We use the \textsc{Pass-8 Fermi} \footnote{Specifically, we employ the \texttt{gll\_iem\_v07} model, available at \url{https://fermi.gsfc.nasa.gov/ssc/data/access/lat/
BackgroundModels.html}} model for the interstellar emission part of the background, which is data-driven, relying less than other models on theoretical assumptions. 
% description of Fermi Bubbles, localised and bright sources if we add the :)...
These background components have been simulated using \texttt{ctools} for a \unit[525]{hour} observation period with the publicly available \texttt{prod3b-v1} IRF library.  The resulting energy spectra are shown in Fig.~\ref{fig:spectra}b.  
The simulated data are then fit using a Gaussian mixture model with ten Gaussians.
The resulting fit is shown in Fig.~\ref{fig:skylocation}b.

\subsection{Marginal likelihoods}

Using the prior distributions for our signal and background models, the next step is to calculate marginal likelihoods.
For each cosmic ray, we calculate two marginal likelihoods, one for the signal hypothesis
\begin{align}\label{eq:LS}
    {\cal L}(d^i|{\cal S}) = \int d\hat\Omega^i 
    \int dE^i \,
    {\cal L}(d^i | \Omega^i E^i) \,
    \pi(\Omega^i, E^i | {\cal S}) ,
\end{align}
and one for the background hypothesis
\begin{align}\label{eq:LB}
    {\cal L}(d^i|{\cal B}) = \int d\hat\Omega^i 
    \int dE^i \,
    {\cal L}(d^i | \Omega^i E^i)
    \pi(\Omega^i, E^i | {\cal B}) .
\end{align}
The marginal likelihoods quantify the support for each hypothesis and the (signal/background) Bayes factor
\begin{align}\label{eq:Bayesfactor}
    \text{BF}^{\mathcal{S},i}_\mathcal{B} = \frac{
    {\cal L}(d^i|{\cal S})
    }{
    {\cal L}(d^i|{\cal B}) 
    } ,
\end{align}
describes the relative likelihood that event $i$ is drawn from one hypothesis versus the other.
However, many events are required to confidently detect a dark-matter annihilation signal.

There are a number of ways to compute the marginal likelihoods.
One option is to represent the likelihoods in Eq.~\ref{eq:likelihood} using \textit{posterior samples}.
These posterior samples are obtained using an uninformative, fiducial prior $\pi_{\text{\o}}$, for example, which is uniform in (the logarithm of) $E$ and isotropic in $\hat\Omega$.
For each event $i$, there are $n_i$ fiducial samples, each consisting of an ordered pair:
\begin{align}
    (E^i_k, \hat\Omega^i_k) .
\end{align}
Here, $k$ is the sample number.
The fiducial samples may be obtained with a stochastic sampler, e.g., \cite{bilby,PyMultiNest,Dynesty,Martinez:2017lzg}.
The fiducial samples can be ``recycled'' (or ``reweighted'') to obtain marginal likelihoods \cite{intro}
\begin{align}\label{eq:recycling}
    {\cal L}(d^i|\mathcal{S}) = & 
    {\cal Z}_{\text{\o}}(d^i) \sum_k^{n_i}
    \frac{\pi(E^i_k, \hat\Omega^i_k | \mathcal{S})}{\pi(E^i_k, \hat\Omega^i_k|\text{\o})} .
\end{align}
Here, $\mathcal{Z}_{\text{\o}}$ is the fiducial evidence obtained with the fiducial prior:
\begin{align}
    {\cal Z}_{\text{\o}}(d^i) = \int d\hat\Omega^i \int dE^i \, 
    {\cal L}(d^i | \hat\Omega^i, E^i) \, 
    \pi(\hat\Omega^i, E^i | \mathcal{S}) .
\end{align}
This procedure---which is a special case of importance sampling \cite{Robert&Casella,Liu}---is convenient because a single set of fiducial samples can be recycled for different analyses (with different signal and/or background distributions)~\footnote{In applied statistics, the fiducial prior (in the denominator of Eq.~\ref{eq:recycling}) is known as the ``proposal distribution'' while the new prior (in the numerator of Eq.~\ref{eq:recycling}) is known as the ``target distribution.''}.

\subsection{Combined likelihood}
The next step is to combine our data to construct a single likelihood function for all the data $\vec{d}$.
If we allow for each event to be either signal or background, then the single-event likelihood can be written as a mixture model of the signal and background hypotheses:
\begin{align}\label{eq:singlelikelihood}
    {\cal L}(d^i| \lambda) = \lambda {\cal L}(d^i | \mathcal{S}) + 
    (1-\lambda) {\cal L}(d^i | \mathcal{B}) .
\end{align}
Here, $\lambda \in (0,1)$ is the probability that event $i$ is drawn from the signal distribution.

Since each measurement is uncorrelated, we can write the combined likelihood as product of single-event likelihoods.
\begin{align}\label{eq:combinedlikelihood}
    {\cal L}(\vec{d} | \lambda) = & \prod_i^N {\cal L}(d^i| \lambda) \nonumber\\
    = & \prod_i^N \lambda {\cal L}(d^i | \mathcal{S}) + 
    (1-\lambda) {\cal L}(d^i | \mathcal{B})
\end{align}
After combining the data for $N$ measurements, the parameter $\lambda$ is interpreted as the proportion of events drawn from the signal hypothesis $(N_{\cal S})$.  
Our goal is to infer $\lambda$, which determines the extent to which a dark-matter signal is present in the data.
The posterior for $\lambda$ is
\begin{align}\label{eq:lambdaposterior}
    p(\lambda | \vec{d}) = \frac{
    {\cal L}(\vec{d} | \lambda) \, \pi(\lambda) 
    }{
    {\cal Z}(\vec{d}) 
    } ,
\end{align}
where $\pi(\lambda)$ is our prior on the mixing fraction and
\begin{align}\label{eq:Z_tot}
    {\cal Z}(\vec{d}) = \int d\lambda \, 
    \pi(\lambda) {\cal L}(\vec{d}|\lambda) ,
\end{align}
is the Bayesian evidence for the data given the full dataset given our signal+background mixture model.
It appears in Eq.~\ref{eq:lambdaposterior} as a normalization constant, but we describe below how it can be used for model selection.
If we take this prior to be uniform \footnote{A more physically motivated prior for $\lambda$ can be obtained by using an $\mathcal{N}_S=0$ Poisson distribution scaled based on the results of previous upper limits on the dark-matter gamma-ray flux.} then the posterior is an order-$N$ polynomial.
However, due to the central limit theorem, the posterior for $\lambda$ converges to a Gaussian distribution as $N$ becomes large and the data become informative.

One can claim a dark-matter detection when this posterior rules out $\lambda=0$ at high credibility.
However, it is desirable to convert statements about $\lambda$ to statements about the velocity-averaged cross section $\langle \sigma v \rangle$.
There are several steps.
First, we relate $\lambda$ to the number of dark-matter annihilation events in the dataset.
Since we expect that the number of background events $(N_{\cal B})$ will be overwhelmingly larger in CTA that the number of signal events, we make the approximation that $N \approx N_{\cal B}$, yielding
\begin{align}\label{eq:lambda}
    \lambda = \frac{N_{\cal S}}{N} \approx \frac{N_{\cal S}}{N_{\cal B}}.
\end{align}

The number of signal events $N_S$ is related to the differential gamma-ray flux from dark-matter annihilation:
\begin{align}\label{eq:NS}
    N_{\cal S} = T \int dE \, \frac{d\Phi}{dE}(E) \, A(E) .
%        N_{\cal S} = T \int \frac{d\Phi}{dE}(E,\psi) A(E) dE .
\end{align}
Here, $T$ is the observation time while $A(E)$ is the energy-dependent collection area of CTA (obtained using \texttt{ctools} \cite{ctools}).
The differential flux $d\Phi/dE$ is linearly related to $\langle \sigma v \rangle$ via Eq.~\ref{eq:flux}.
Thus, there is a one-to-one mapping between $\lambda$ and $\langle \sigma v \rangle$.

\section{Demonstration}\label{demo}
In this section we demonstrate our formalism using mock data targeting the Galactic Centre.
We simulate both astrophysical background and dark-matter annihilation signal assuming a scalar singlet model.
We generate $\unit[525]{hours}$ ($\unit[21]{days}$) of background data using random draws from the background distributions described in Section~\ref{formalism}.  The background distribution of sky location is shown in Fig.~\ref{fig:skylocation}b while the background energy spectra is shown in Fig.~\ref{fig:spectra}b.
This observation time corresponds to what we might expect to obtain for CTA measurements of the galactic centre \cite{CTAprecon}.
This dataset corresponds to $\approx 10^8$ gamma-ray events.
To this dataset we swap in a small number of signal events.  The signal distribution of sky location is shown in Fig.~\ref{fig:skylocation}a while the signal energy spectra is shown in Fig.~\ref{fig:spectra}a.
We vary the precise number of signal events in order to control the statistical significance of the dark-matter annihilation signal.

\begin{figure*}[htpb!]
     \centering
     \begin{subfigure}[b]{0.49\textwidth}
         \centering \includegraphics[width=\textwidth]{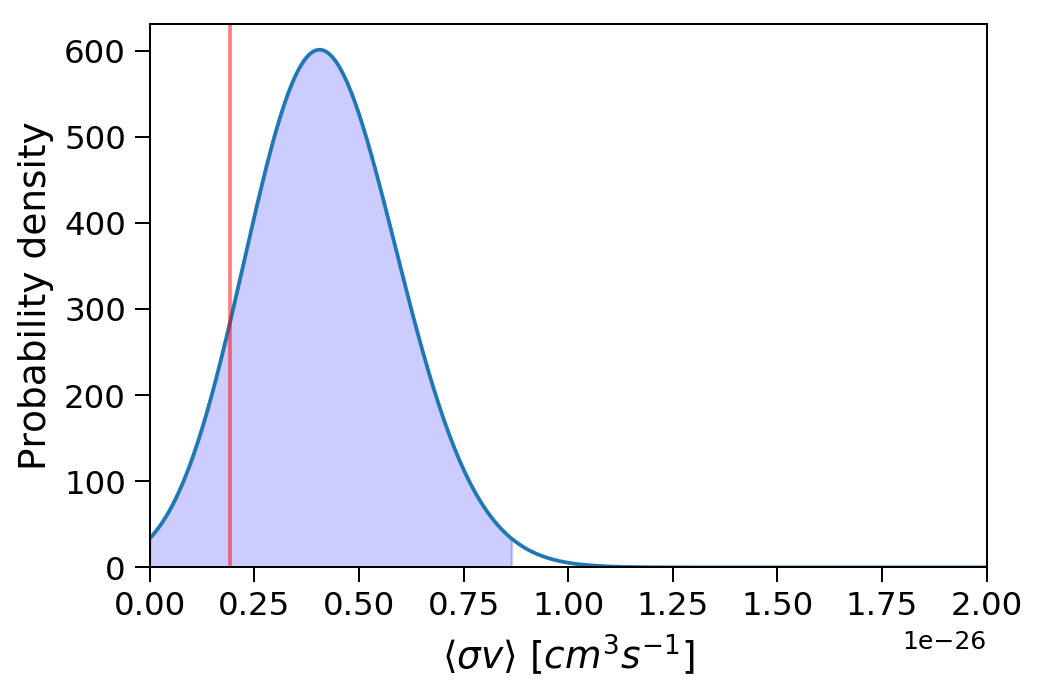}
         \caption{}
     \end{subfigure}
     \hfill
     \begin{subfigure}[b]{0.49\textwidth}
         \centering \includegraphics[width=\textwidth]{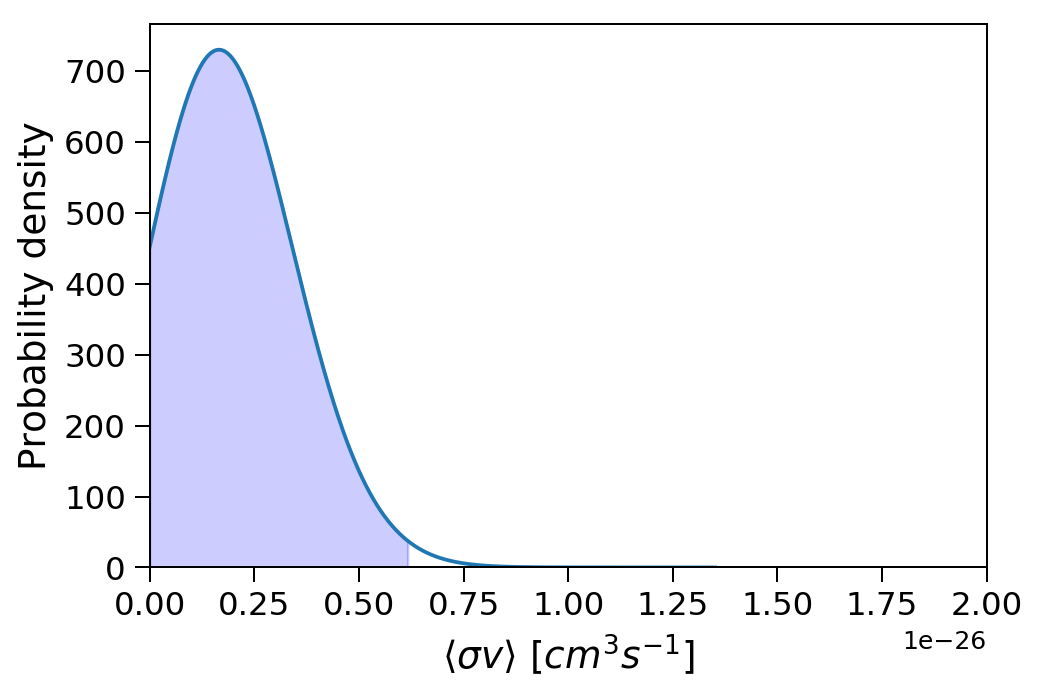}
         \caption{}
     \end{subfigure}
        \caption{Posterior plots of $\langle\sigma v \rangle$ for the scalar singlet model, $m_{\chi} = \unit[2]{TeV}$ with a 99$\%$ credible interval indicated by the shaded region. \textit{Left}: a marginal detection at 99\% credibility. The true value of $\langle \sigma v \rangle$ is indicated by the vertical red line.
        \textit{Right}: a non-detection yields an upper limit of $\langle\sigma v \rangle_{\textrm{Upper Lim}} = \unit[6.17\times 10^{-27}]{cm^3s^{-1}}$ (99\% credibility).}
        \label{fig:scalarsinglet}
\end{figure*}

First, we tune the number of signal events to illustrate what a dark-matter detection looks like in our Bayesian formalism.
In Fig.~\ref{fig:scalarsinglet}a, we plot the posterior distribution for $\langle \sigma v \rangle$ using a dataset for which we can (only just) rule out the null hypothesis $\langle \sigma v \rangle=0$ with 99\% credibility. In this case we set the mass of the dark matter particle to $m_{\chi} = \unit[2]{TeV}$.
In order to achieve this statistical significance, we employed $7.6\times 10^4$ signal events out of a total set of $1.08\times 10^8$ events ($0.07\%$).
The vertical red line in Fig.~\ref{fig:scalarsinglet}a indicates the true value of $\langle \sigma v \rangle$.
The posterior is statistically consistent with the true value, though, it does not peak at the true value of $\langle \sigma v \rangle$ due to statistical fluctuations.
The shaded region shows the (highest posterior density) 99\% credible interval, which covers $\unit[(0,8.64)\times 10^{-27}]{cm^{-3}s^{-1}}$. 

Next, we set the number of signal events to zero in order to illustrate what a non-detection result looks like.
In Fig.~\ref{fig:scalarsinglet}b, we plot the posterior distribution for $\langle \sigma v \rangle$, again setting $m_{\chi} = \unit[2]{TeV}$.
In this panel, we cannot rule out $\langle \sigma v \rangle=0$ with even marginal significance, and so we obtain only an upper limit of $\langle \sigma v \rangle > \unit[6.17\times 10^{-27}]{cm^{-3}s^{-1}}$ (99\%) credibility.
As the observation time $T$ is increased, the typical upper limit scales like $T^{-1/2}$. \footnote{In order to reduce computation time, we have employed a total set of $1.08\times 10^5$ events and scaled the upper limit using this well known scaling relation to present the expected upper limits for $10^8$ events.}

\begin{figure*}[htpb!]
\includegraphics[width = \textwidth]{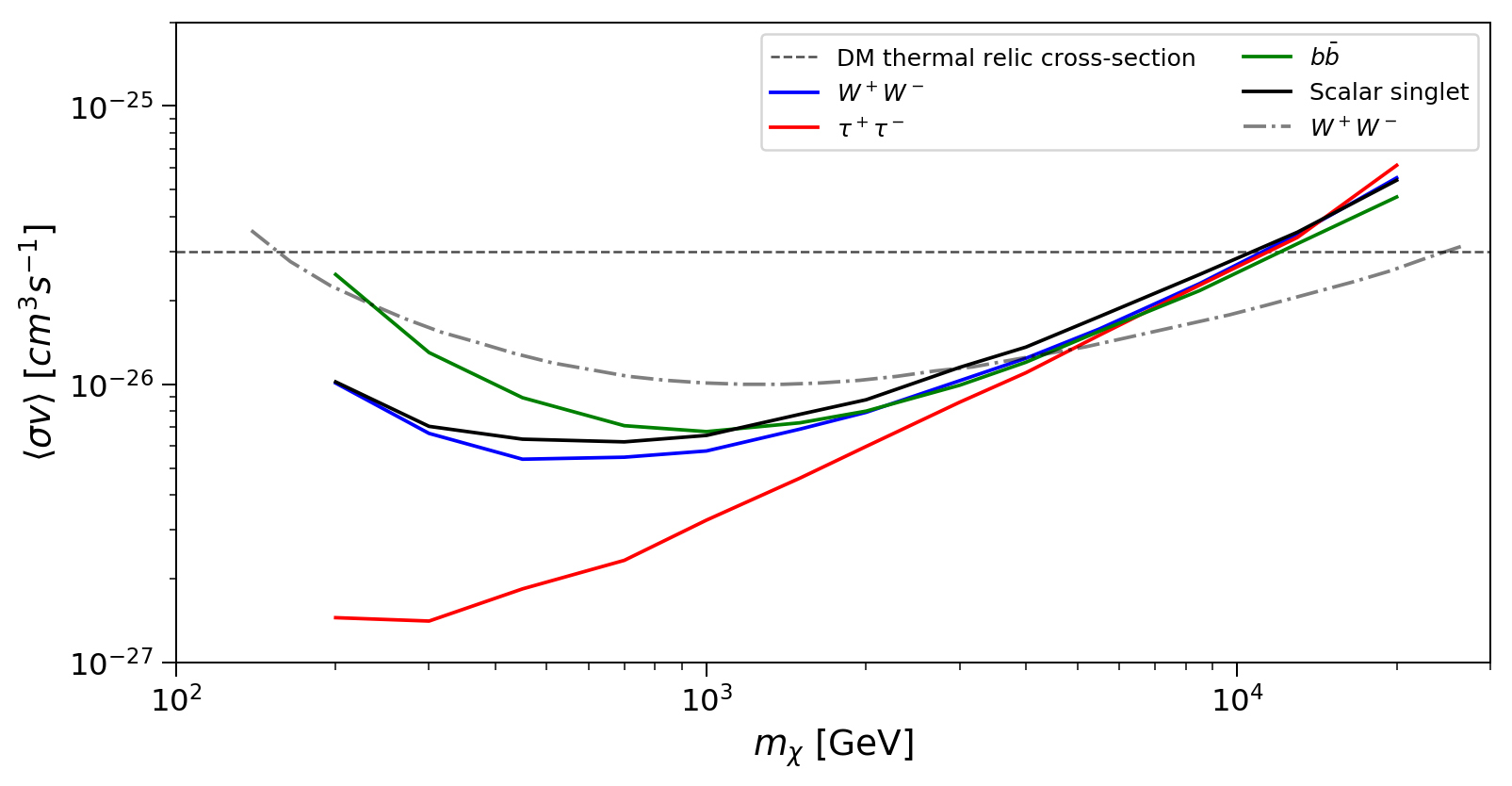}
\caption{Projected 99\% upper limits on the velocity-averaged cross section $\langle \sigma v \rangle$ as a function of dark-matter mass $m_\chi$. The limits are calculated assuming $\unit[525]{hours}$ of observation of the Galactic Centre. We assume simple models whereby the dark matter particle annihilates completely to $W^+W^-$ (with electroweak corrections) (blue), $\tau^+\tau^-$ (red) and $b\bar{b}$ (green), and the scalar singlet model (black). The dash-dotted $W^+W^-$ line is reproduced from Ref. \cite{CTAprecon}.
The dashed horizontal line corresponds to the dark matter `thermal' cross-section (Eq. \ref{eq:thermalcross}).
The projected upper limits fall below the dashed black line for dark-matter masses of $\approx \unit[0.2-5]{TeV}$, indicating that CTA has the required sensitivity to detect dark matter in this range.}
\label{fig:exclusioncurves}
\end{figure*}

In order to show how our results vary with the dark-matter mass, we calculate the Bayesian upper limit (99\% credibility) for $\langle \sigma v \rangle$ (assuming null results) for different values of $m_\chi$.
By changing $m_\chi$, we change the expected signal distribution---$\pi(E|\mathcal{S})$ (Eq.~\ref{eq:pi(E|S)})---which, in turn, affects the extent to which signal can be distinguished from background.
For each value of $m_\chi$, we calculate the 99\% upper limit for the case where no signal events are present in the data.
In Fig.~\ref{fig:exclusioncurves}, we plot these limits on $\langle \sigma v \rangle$ as a function of $m_\chi$.  The coloured curves represent different simplified models of dark matter annihilation final state, and the scalar singlet dark matter model.  

We reproduced the $W^+W^-$ limit, shown by the dash-dotted line, from Ref. \cite{CTAprecon}.  When comparing the two $W^+W^-$ limits we have to keep in mind that, among various differences, they are extracted by different statistical methods and are based on a different observational strategy.  Our limit, for example, relies on the full dataset arriving from the Galactic Centre while the one in Ref. \cite{CTAprecon} is based on an On-Off observing strategy.  The latter reduces the number of detected photons from dark matter annihilation and is expected to produce a slightly weaker limit.  We account the difference in shape to other differences in the inference such as the different statistical meaning of frequentist confidence and Bayesian credibility intervals. 

\section{Discussion}\label{discussion}
Bayesian methods are well known, rigorous statistical tools which are widely used in (particle) astrophysics.  Bayesian inference is useful because of its transparency, namely, that it can simply and quantify various assumptions and their effect on the inferred quantities.  Due to the popularity of Bayesian methods various numerical tools, such as likelihood calculators or samplers, exist that (future extensions of) our analysis can take advantage of.  Because of this, our Bayesian analysis is flexible and expandable, since it is easy to change or add models, assumptions, and uncertainties without altering the statistical framework itself.  Additionally, with small extensions, the same framework can be used for multiple purposes such as parameter estimation or model comparison. 
The analysis presented here, for example, can be extended to take advantage of the full machinery of Bayesian inference.
We sketch out some of the most important next steps in order to describe how this formalism can be used (i) to measure model parameters, (ii) to take into account systematic error, and (iii) for Bayesian model selection. 

\textit{Measuring model parameters.}---In the demonstration above, we fix the dark matter mass $m_\chi$ to just one value at a time.
However, in practice, the dark-matter mass should be treated as a free parameter.
Some dark-matter models have multiple parameters, as is the case with the scalar singlet model described in Section~\ref{formalism}; Eq.~\ref{eq:scalarsinglet}, which describes this model, includes three independent parameters.
We denote the list of parameters associated with a model $\theta$ so that, for example, $\theta = \{J, m_\chi, \lambda_{Sh}, \lambda_S\}$.
The signal prior is conditional on these parameters \footnote{In this discussion, we sometimes denote the signal or background labels $\mathcal{S}$ and $\mathcal{B}$ as subscripts for readability.}:
\begin{align}\label{eq:hierarchical}
    \pi_{\mathcal{S}}(\hat\Omega, E | \theta) .
\end{align}
The parameter $\theta$ is sometimes referred to as a \textit{hyper-parameter} \cite{intro} since it controls the shape of the distribution of other parameters $(E,\hat\Omega)$.
Models like Eq.~\ref{eq:hierarchical}, where the prior is conditional on one or more hyper-parameters, are referred to as \textit{hierarchical} models \cite{intro}.
It is necessary to marginalize over the free parameters in $\theta$ when calculating the marginal signal likelihood.
The generalization of Eq.~\ref{eq:LS} is 
\begin{widetext}
\begin{align}
    {\cal L}_\mathcal{S}(d^i | \theta) = 
    \int d\hat\Omega^i
    \int dE^i \,
    {\cal L}(d^i | \hat\Omega^i, E^i) \,
    \pi_{\mathcal{S}}(\hat\Omega^i, E^i | \theta) \, 
    \pi(\theta) .
\end{align}
\end{widetext}

The recycling trick described in Eq.~\ref{eq:recycling} is useful here since we can use one set of fiducial samples for many different values of $\theta$.  The likelihood of the complete data given signal fraction $\lambda$  and dark-matter parameters $\theta$ is
\begin{widetext}
\begin{align}\label{eq:full_likelihood}
    {\cal L}(\vec{d} | \lambda, \theta) = \prod_i^N
    \left(
    (1-\lambda) {\cal L}(d^i | \mathcal{B}) + 
    \lambda {\cal Z}_{\text{\o}}(d^i)
    \sum_{k=1}^{n_i} \frac{\pi_{\mathcal{S}}(E^i_k, \hat\Omega^i_k | \theta)}
    {\pi(E^i_k, \hat\Omega^i_k | \text{\o})}
    \right) .
\end{align}
\end{widetext}
Programming the likelihood into a stochastic sampler, we can obtain posterior samples for $\theta$, allowing us to, for example, to measure the dark-matter mass in case of signal detection.
The sum over posterior samples can sometimes be evaluated rapidly using graphical processor units (GPUs) \cite{gw_population}.

\textit{Taking into account systematic error.}---In our demonstration, we have assumed that the astrophysical background model is perfectly well specified.
In reality, the flux of astrophysical cosmic rays is an active area of research, and so our model is subject to uncertainty (discussed in Section~\ref{formalism}D).
One way to accommodate this systematic uncertainty in our framework is with noise-model hyper-parameters, which we denote by $\eta$:
\begin{align}
    \pi_\mathcal{B}(\hat\Omega| \eta) .
\end{align}
The power-law slope of the Einasto profile, for example, can be treated as a free parameter.  If we choose, we can add this slope to $\eta$.
It is necessary to marginalize over the free parameters in $\theta$ when calculating the marginal signal likelihood.
The generalization of Eq.~\ref{eq:LB} is 
\begin{widetext}
\begin{align}
    {\cal L}_{\mathcal{B}}(d^i | \eta) = 
    \int d\hat\Omega^i
    \int dE^i \,
    {\cal L}(d^i | \hat\Omega^i, E^i) \,
    \pi_{\mathcal{B}}(\hat\Omega^i | \eta) .
\end{align}
\end{widetext}
As in Eq.~\ref{eq:full_likelihood}, one may recycle fiducial background samples for computational efficiency.

\textit{Model selection.}---Above, we carried out our demonstration with a single dark-matter model.
However, it will likely be useful to perform analyses for multiple dark-matter models in order to see which one best fits the data.
In order to do this, one may calculate the Bayesian evidence for the entire dataset (as in Eq.~\ref{eq:Z_tot}) given each dark-matter model $(\mathcal{S}_1, \mathcal{S}_2, ...)$:
\begin{align}
    {\cal Z}(\vec{d} | \alpha) = 
    \int d\lambda \, \pi(\lambda) \,
    {\cal L}(\vec{d} | \lambda, \alpha) .
\end{align}
Here, $\alpha$ is an index that numbers the different dark-matter models.
The Bayes factor comparing models $\alpha=1$ and $\alpha=2$ is
\begin{align}
    \text{BF}^2_1 = \frac{
    {\cal Z}(\vec{d} | \alpha=2)
    }{
    {\cal Z}(\vec{d} | \alpha=1)
    } .
\end{align}
A large Bayes factor (for example, $\ln\text{BF}>8$) indicates that one dark-matter model is strongly preferred to another \cite{intro}.
One should compare the Bayes factor to the ratio of maximum likelihood values for each model in order to understand the extent to which the Bayes factor is influenced by the prior distributions of each model:
\begin{align}
    \frac
    {\max_{\theta_2} {\cal L}(\vec{d}|\theta_2, \alpha=2)}
    {\max_{\theta_1} {\cal L}(\vec{d} | \theta_1, \alpha=1)} .
\end{align}
Here, $\theta_1, \theta_2$ are the parameters for model $\alpha=1,2$ respectively.

Indirect searches for dark matter via gamma rays considerably matured during the last decade.  Observations of the Fermi Gamma-ray Space Telescope \cite{Fermi-LAT:2009ihh} and those of H.E.S.S. \cite{HESS:2018cbt}, among others, paved the way toward precision measurements of cosmic gamma rays.  The capabilities of CTA will surpass those of the earlier gamma ray instruments on the front of energy range, resolution and sensitivity.  This opens an exciting new era in the next decade of dark matter indirect detection.  The increasingly precise gamma-ray observations have to be matched by increasingly sophisticated inference tools to determine the fundamental physical properties of dark matter particles, such as their mass, spin and interaction strengths.  In this spirit, the formalism presented in this work provides one of the potentially useful frameworks for CTA measurements of dark matter.

\section{Acknowledgements}
We are grateful to Christopher Eckner and Torsten Bringmann for collaboration on the early version of this project based on frequentist inference.  We also thank Giacomo Damico, Fabio Iocco, Manuel Meyer, and Manuela Vecchi for valuable comments on the manuscript.  ET is supported by the Australian Research Council (ARC) Centre of Excellence CE170100004.  The work of CB is supported by the ARC Discovery Project grants DP180102209 and DP210101636.

\bibliographystyle{JHEP}
\bibliography{Bibliography}

\providecommand{\href}[2]{#2}\begingroup\raggedright\begin{thebibliography}{10}

\bibitem{cta}
{\scshape CTA Consortium} collaboration, B.S.~Acharya et~al., \emph{{Science
  with the Cherenkov Telescope Array}}, WSP (11, 2018),
  \href{https://doi.org/10.1142/10986}{10.1142/10986},
  [\href{https://arxiv.org/abs/1709.07997}{{\ttfamily 1709.07997}}].

\bibitem{bertone}
G.~Bertone, D.~Hooper and J.~Silk, \emph{Particle dark matter: evidence,
  candidates and constraints},
  \href{https://doi.org/10.1016/j.physrep.2004.08.031}{\emph{Physics Reports}
  {\bfseries 405} (2005) 279–390}.

\bibitem{berg}
L.~Bergström, \emph{Dark matter candidates},
  \href{https://doi.org/10.1088/1367-2630/11/10/105006}{\emph{New Journal of
  Physics} {\bfseries 11} (2009) 105006}.

\bibitem{GAMBIT:2021rlp}
{\scshape GAMBIT} collaboration, \emph{{Thermal WIMPs and the scale of new
  physics: global fits of Dirac dark matter effective field theories}},
  \href{https://doi.org/10.1140/epjc/s10052-021-09712-6}{\emph{Eur. Phys. J. C}
  {\bfseries 81} (2021) 992}
  [\href{https://arxiv.org/abs/2106.02056}{{\ttfamily 2106.02056}}].

\bibitem{GAMBIT:2018gjo}
{\scshape GAMBIT} collaboration, \emph{{Combined collider constraints on
  neutralinos and charginos}},
  \href{https://doi.org/10.1140/epjc/s10052-019-6837-x}{\emph{Eur. Phys. J. C}
  {\bfseries 79} (2019) 395}
  [\href{https://arxiv.org/abs/1809.02097}{{\ttfamily 1809.02097}}].

\bibitem{GAMBIT:2018eea}
{\scshape GAMBIT} collaboration, \emph{{Global analyses of Higgs portal singlet
  dark matter models using GAMBIT}},
  \href{https://doi.org/10.1140/epjc/s10052-018-6513-6}{\emph{Eur. Phys. J. C}
  {\bfseries 79} (2019) 38} [\href{https://arxiv.org/abs/1808.10465}{{\ttfamily
  1808.10465}}].

\bibitem{Athron:2018ipf}
P.~Athron, J.M.~Cornell, F.~Kahlhoefer, J.~Mckay, P.~Scott and S.~Wild,
  \emph{{Impact of vacuum stability, perturbativity and XENON1T on global fits
  of $\mathbb {Z}_2$ and $\mathbb {Z}_3$ scalar singlet dark matter}},
  \href{https://doi.org/10.1140/epjc/s10052-018-6314-y}{\emph{Eur. Phys. J. C}
  {\bfseries 78} (2018) 830}
  [\href{https://arxiv.org/abs/1806.11281}{{\ttfamily 1806.11281}}].

\bibitem{GAMBIT:2017snp}
{\scshape GAMBIT} collaboration, \emph{{Global fits of GUT-scale SUSY models
  with GAMBIT}},
  \href{https://doi.org/10.1140/epjc/s10052-017-5167-0}{\emph{Eur. Phys. J. C}
  {\bfseries 77} (2017) 824}
  [\href{https://arxiv.org/abs/1705.07935}{{\ttfamily 1705.07935}}].

\bibitem{GAMBIT:2017zdo}
{\scshape GAMBIT} collaboration, \emph{{A global fit of the MSSM with GAMBIT}},
  \href{https://doi.org/10.1140/epjc/s10052-017-5196-8}{\emph{Eur. Phys. J. C}
  {\bfseries 77} (2017) 879}
  [\href{https://arxiv.org/abs/1705.07917}{{\ttfamily 1705.07917}}].

\bibitem{GAMBIT:2017gge}
{\scshape GAMBIT} collaboration, \emph{{Status of the scalar singlet dark
  matter model}},
  \href{https://doi.org/10.1140/epjc/s10052-017-5113-1}{\emph{Eur. Phys. J. C}
  {\bfseries 77} (2017) 568}
  [\href{https://arxiv.org/abs/1705.07931}{{\ttfamily 1705.07931}}].

\bibitem{buckley}
M.R.~Buckley and A.H.~Peter, \emph{Gravitational probes of dark matter
  physics}, \href{https://doi.org/10.1016/j.physrep.2018.07.003}{\emph{Physics
  Reports} {\bfseries 761} (2018) 1–60}.

\bibitem{Bernabei2008}
R.~Bernabei et~al., \emph{{First results from DAMA/LIBRA and the combined
  results with DAMA/NaI}}, {\emph{Euro. Phys. J. C} {\bfseries 56} (2008) 333}.

\bibitem{ros}
L.~Roszkowski, E.M.~Sessolo and S.~Trojanowski, \emph{Wimp dark matter
  candidates and searches—current status and future prospects},
  \href{https://doi.org/10.1088/1361-6633/aab913}{\emph{Reports on Progress in
  Physics} {\bfseries 81} (2018) 066201}.

\bibitem{einasto}
L.~Pieri, J.~Lavalle, G.~Bertone and E.~Branchini, \emph{Implications of
  high-resolution simulations on indirect dark matter searches},
  \href{https://doi.org/10.1103/PhysRevD.83.023518}{\emph{Phys. Rev. D}
  {\bfseries 83} (2011) 023518}.

\bibitem{viana}
A.~Viana, H.~Schoorlemmer, A.~Albert, V.~de~Souza, J.P.~Harding and J.~Hinton,
  \emph{Searching for dark matter in the galactic halo with a wide field of
  view tev gamma-ray observatory in the southern hemisphere},
  \href{https://doi.org/10.1088/1475-7516/2019/12/061}{\emph{Journal of
  Cosmology and Astroparticle Physics} {\bfseries 2019} (2019) 061–061}.

\bibitem{Catena}
R.~Catena and P.~Ullio, \emph{{A novel determination of the local dark matter
  density}}, \href{https://doi.org/10.1088/1475-7516/2010/08/004}{\emph{JCAP}
  {\bfseries 08} (2010) 004} [\href{https://arxiv.org/abs/0907.0018}{{\ttfamily
  0907.0018}}].

\bibitem{CTAprecon}
A.~Acharyya, R.~Adam, C.~Adams, I.~Agudo, A.~Aguirre-Santaella, R.~Alfaro
  et~al., \emph{Sensitivity of the cherenkov telescope array to a dark matter
  signal from the galactic centre},
  \href{https://doi.org/10.1088/1475-7516/2021/01/057}{\emph{Journal of
  Cosmology and Astroparticle Physics} {\bfseries 2021} (2021) 057–057}.

\bibitem{Lefranc}
V.~Lefranc, E.~Moulin, P.~Panci and J.~Silk, \emph{Prospects for annihilating
  dark matter in the inner galactic halo by the cherenkov telescope array},
  \href{https://doi.org/10.1103/PhysRevD.91.122003}{\emph{Phys. Rev. D}
  {\bfseries 91} (2015) 122003}.

\bibitem{Hutten}
M.~H\"utten, C.~Combet, G.~Maier and D.~Maurin, \emph{{Dark matter substructure
  modelling and sensitivity of the Cherenkov Telescope Array to Galactic dark
  halos}}, \href{https://doi.org/10.1088/1475-7516/2016/09/047}{\emph{JCAP}
  {\bfseries 09} (2016) 047}
  [\href{https://arxiv.org/abs/1606.04898}{{\ttfamily 1606.04898}}].

\bibitem{Pierre}
M.~Pierre, J.M.~Siegal-Gaskins and P.~Scott, \emph{{Sensitivity of CTA to dark
  matter signals from the Galactic Center}},
  \href{https://doi.org/10.1088/1475-7516/2014/10/E01}{\emph{JCAP} {\bfseries
  06} (2014) 024} [\href{https://arxiv.org/abs/1401.7330}{{\ttfamily
  1401.7330}}].

\bibitem{Silverwood}
H.~Silverwood, C.~Weniger, P.~Scott and G.~Bertone, \emph{{A realistic
  assessment of the CTA sensitivity to dark matter annihilation}},
  \href{https://doi.org/10.1088/1475-7516/2015/03/055}{\emph{JCAP} {\bfseries
  03} (2015) 055} [\href{https://arxiv.org/abs/1408.4131}{{\ttfamily
  1408.4131}}].

\bibitem{Aguirre-Santaella}
A.~Aguirre-Santaella, V.~Gammaldi, M.A.~S\'anchez-Conde and D.~Nieto,
  \emph{{Cherenkov Telescope Array sensitivity to branon dark matter models}},
  \href{https://doi.org/10.1088/1475-7516/2020/10/041}{\emph{JCAP} {\bfseries
  10} (2020) 041} [\href{https://arxiv.org/abs/2006.16706}{{\ttfamily
  2006.16706}}].

\bibitem{intro}
E.~Thrane and C.~Talbot, \emph{{An introduction to Bayesian inference in
  gravitational-wave astronomy: parameter estimation, model selection, and
  hierarchical models}}, {\emph{Pub. Astron. Soc. Aust.} {\bfseries 36} (2019)
  E010}.

\bibitem{ctools}
J.~Knödlseder, L.~Tibaldo, D.~Tiziani, A.~Specovius, J.~Cardenzana, M.~Mayer
  et~al., ``ctools 1.7.4.'' Available at
  \url{https://doi.org/10.5281/zenodo.4727876}, Apr., 2021.
\newblock 10.5281/zenodo.4727876.

\bibitem{sim_edisp}
J.~Knödlseder, L.~Tibaldo, D.~Tiziani, A.~Specovius, J.~Cardenzana, M.~Mayer
  et~al., ``ctools 1.7.4.'' Available at
  \url{https://github.com/gammalib/gammalib/blob/ad26baa0c58c305a05fe718f66c84326bd93699e/inst/cta/test/dev/test_sim_edisp.py},
  Apr., 2021.
\newblock 10.5281/zenodo.4727876.

\bibitem{sim_psf}
J.~Knödlseder, L.~Tibaldo, D.~Tiziani, A.~Specovius, J.~Cardenzana, M.~Mayer
  et~al., ``ctools 1.7.4.'' Available at
  \url{https://github.com/gammalib/gammalib/blob/ad26baa0c58c305a05fe718f66c84326bd93699e/inst/cta/test/dev/test_sim_psf.py},
  Apr., 2021.
\newblock 10.5281/zenodo.4727876.

\bibitem{holz}
D.~Holz and A.~Zee, \emph{Collisional dark matter and scalar phantoms},
  \href{https://doi.org/https://doi.org/10.1016/S0370-2693(01)01033-4}{\emph{Physics
  Letters B} {\bfseries 517} (2001) 239}.

\bibitem{burgess}
C.P.~Burgess, M.~Pospelov and T.~ter Veldhuis, \emph{{The Minimal model of
  nonbaryonic dark matter: A Singlet scalar}},
  \href{https://doi.org/10.1016/S0550-3213(01)00513-2}{\emph{Nucl. Phys. B}
  {\bfseries 619} (2001) 709}
  [\href{https://arxiv.org/abs/hep-ph/0011335}{{\ttfamily hep-ph/0011335}}].

\bibitem{McDonald}
J.~McDonald, \emph{{Gauge singlet scalars as cold dark matter}},
  \href{https://doi.org/10.1103/PhysRevD.50.3637}{\emph{Phys. Rev. D}
  {\bfseries 50} (1994) 3637}
  [\href{https://arxiv.org/abs/hep-ph/0702143}{{\ttfamily hep-ph/0702143}}].

\bibitem{Yaguna}
C.E.~Yaguna, \emph{{Gamma rays from the annihilation of singlet scalar dark
  matter}}, \href{https://doi.org/10.1088/1475-7516/2009/03/003}{\emph{JCAP}
  {\bfseries 03} (2009) 003} [\href{https://arxiv.org/abs/0810.4267}{{\ttfamily
  0810.4267}}].

\bibitem{Profumo}
S.~Profumo, L.~Ubaldi and C.~Wainwright, \emph{{Singlet Scalar Dark Matter:
  monochromatic gamma rays and metastable vacua}},
  \href{https://doi.org/10.1103/PhysRevD.82.123514}{\emph{Phys. Rev. D}
  {\bfseries 82} (2010) 123514}
  [\href{https://arxiv.org/abs/1009.5377}{{\ttfamily 1009.5377}}].

\bibitem{Arina}
C.~Arina and M.H.G.~Tytgat, \emph{{Constraints on Light WIMP candidates from
  the Isotropic Diffuse Gamma-Ray Emission}},
  \href{https://doi.org/10.1088/1475-7516/2011/01/011}{\emph{JCAP} {\bfseries
  01} (2011) 011} [\href{https://arxiv.org/abs/1007.2765}{{\ttfamily
  1007.2765}}].

\bibitem{Mambrini}
Y.~Mambrini, \emph{{Higgs searches and singlet scalar dark matter: Combined
  constraints from XENON 100 and the LHC}},
  \href{https://doi.org/10.1103/PhysRevD.84.115017}{\emph{Phys. Rev. D}
  {\bfseries 84} (2011) 115017}
  [\href{https://arxiv.org/abs/1108.0671}{{\ttfamily 1108.0671}}].

\bibitem{micromegas}
G.~Belanger, F.~Boudjema and A.~Pukhov, \emph{micromegas : a code for the
  calculation of dark matter properties in generic models of particle
  interaction},  2014.

\bibitem{PPPC}
M.~Cirelli, G.~Corcella, A.~Hektor, G.~Hütsi, M.~Kadastik, P.~Panci et~al.,
  \emph{Pppc 4 dm id: a poor particle physicist cookbook for dark matter
  indirect detection},
  \href{https://doi.org/10.1088/1475-7516/2011/03/051}{\emph{Journal of
  Cosmology and Astroparticle Physics} {\bfseries 2011} (2011) 051–051}.

\bibitem{Burkert}
A.~Burkert, \emph{The structure of dark matter halos in dwarf galaxies},
  \href{https://doi.org/10.1086/309560}{\emph{The Astrophysical Journal}
  {\bfseries 447} (1995) }.

\bibitem{Iocco:2016itg}
F.~Iocco and M.~Benito, \emph{{An estimate of the DM profile in the Galactic
  bulge region}}, \href{https://doi.org/10.1016/j.dark.2016.12.004}{\emph{Phys.
  Dark Univ.} {\bfseries 15} (2017) 90}
  [\href{https://arxiv.org/abs/1611.09861}{{\ttfamily 1611.09861}}].

\bibitem{Benito:2016kyp}
M.~Benito, N.~Bernal, N.~Bozorgnia, F.~Calore and F.~Iocco, \emph{{Particle
  Dark Matter Constraints: the Effect of Galactic Uncertainties}},
  \href{https://doi.org/10.1088/1475-7516/2017/02/007}{\emph{JCAP} {\bfseries
  02} (2017) 007} [\href{https://arxiv.org/abs/1612.02010}{{\ttfamily
  1612.02010}}].

\bibitem{Heros}
C.~P\'erez de~los Heros, \emph{{Status of direct and indirect dark matter
  searches}}, \href{https://doi.org/10.22323/1.364.0694}{\emph{PoS} {\bfseries
  EPS-HEP2019} (2020) 694} [\href{https://arxiv.org/abs/2001.06193}{{\ttfamily
  2001.06193}}].

\bibitem{benito}
M.~Benito, F.~Iocco and A.~Cuoco, \emph{Uncertainties in the galactic dark
  matter distribution: An update},
  \href{https://doi.org/10.1016/j.dark.2021.100826}{\emph{Physics of the Dark
  Universe} {\bfseries 32} (2021) 100826}.

\bibitem{Blandford}
R.~Blandford and D.~Eichler, \emph{Particle acceleration at astrophysical
  shocks: A theory of cosmic ray origin},
  \href{https://doi.org/https://doi.org/10.1016/0370-1573(87)90134-7}{\emph{Physics
  Reports} {\bfseries 154} (1987) 1}.

\bibitem{anjos}
R.C.~Anjos and F.~Catalani, \emph{Galactic center as an efficient source of
  cosmic rays},
  \href{https://doi.org/10.1103/physrevd.101.123015}{\emph{Physical Review D}
  {\bfseries 101} (2020) }.

\bibitem{reynolds}
S.~Reynolds, \emph{Supernova remnants at high energy.},
  \href{https://doi.org/10.1146/annurev.astro.46.060407.145237}{\emph{Annual
  Review of Astronomy and Astrophysics} {\bfseries 46} (2008) 89}.

\bibitem{fujita}
Y.~Fujita, K.~Murase and S.S.~Kimura, \emph{Sagittarius a* as an origin of the
  galactic pev cosmic rays?},
  \href{https://doi.org/10.1088/1475-7516/2017/04/037}{\emph{Journal of
  Cosmology and Astroparticle Physics} {\bfseries 2017} (2017) 037–037}.

\bibitem{HESS}
H.~Abdalla, A.~Abramowski, F.~Aharonian, F.~Ait~Benkhali, E.O.~Angüner,
  M.~Arakawa et~al., \emph{The h.e.s.s. galactic plane survey},
  \href{https://doi.org/10.1051/0004-6361/201732098}{\emph{Astronomy \&
  Astrophysics} {\bfseries 612} (2018) A1}.

\bibitem{guepin}
C.~Guépin, L.~Rinchiuso, K.~Kotera, E.~Moulin, T.~Pierog and J.~Silk,
  \emph{Pevatron at the galactic center: multi-wavelength signatures from
  millisecond pulsars},
  \href{https://doi.org/10.1088/1475-7516/2018/07/042}{\emph{Journal of
  Cosmology and Astroparticle Physics} {\bfseries 2018} (2018) 042–042}.

\bibitem{cheng}
K.-S.~Cheng, D.O.~Chernyshov, V.A.~Dogiel, C.-M.~Ko, W.-H.~Ip and Y.~Wang,
  \emph{The fermi bubble as a source of cosmic rays in the energy range
  >1015ev}, \href{https://doi.org/10.1088/0004-637x/746/2/116}{\emph{The
  Astrophysical Journal} {\bfseries 746} (2012) 116}.

\bibitem{bilby}
G.~Ashton et~al., \emph{{Bilby: A user-friendly Bayesian inference library for
  gravitational-wave astronomy}}, {\emph{Astrophys. J. Supp.} {\bfseries 241}
  (2019) 27}.

\bibitem{PyMultiNest}
J.~{Buchner}, A.~{Georgakakis}, K.~{Nandra}, L.~{Hsu}, C.~{Rangel},
  M.~{Brightman} et~al., \emph{{X-ray spectral modelling of the AGN obscuring
  region in the CDFS: Bayesian model selection and catalogue}},
  \href{https://doi.org/10.1051/0004-6361/201322971}{\emph{Astron. Astrophys.}
  {\bfseries 564} (2014) A125}
  [\href{https://arxiv.org/abs/1402.0004}{{\ttfamily 1402.0004}}].

\bibitem{Dynesty}
J.S.~{Speagle}, \emph{{DYNESTY: a dynamic nested sampling package for
  estimating Bayesian posteriors and evidences}},
  \href{https://doi.org/10.1093/mnras/staa278}{\emph{Mon. Not. R. Ast. Soc.
  Lett.} {\bfseries 493} (2020) 3132}
  [\href{https://arxiv.org/abs/1904.02180}{{\ttfamily 1904.02180}}].

\bibitem{Martinez:2017lzg}
{\scshape GAMBIT} collaboration, \emph{{Comparison of statistical sampling
  methods with ScannerBit, the GAMBIT scanning module}},
  \href{https://doi.org/10.1140/epjc/s10052-017-5274-y}{\emph{Eur. Phys. J. C}
  {\bfseries 77} (2017) 761}
  [\href{https://arxiv.org/abs/1705.07959}{{\ttfamily 1705.07959}}].

\bibitem{Robert&Casella}
C.P.~Robert and G.~Casella, \emph{Monte Carlo Statistical Methods}, Springer,
  New York, 2~ed. (2004).

\bibitem{Liu}
J.S.~Liu, \emph{Monte Carlo Strategies in Scientific Computing}, Springer, New
  York, 1~ed. (2004).

\bibitem{gw_population}
{\bf C. Talbot}, {\bf R. J. E. Smith}, E.~Thrane and G.B.~Poole,
  \emph{{Parallelized Inference for Gravitational-Wave Astronomy }},
  {\emph{Phys. Rev. D} {\bfseries 100} (2019) 043030}.

\bibitem{Fermi-LAT:2009ihh}
{\scshape Fermi-LAT} collaboration, \emph{{The Large Area Telescope on the
  Fermi Gamma-ray Space Telescope Mission}},
  \href{https://doi.org/10.1088/0004-637X/697/2/1071}{\emph{Astrophys. J.}
  {\bfseries 697} (2009) 1071}
  [\href{https://arxiv.org/abs/0902.1089}{{\ttfamily 0902.1089}}].

\bibitem{HESS:2018cbt}
{\scshape HESS} collaboration, \emph{{Search for $\gamma$-Ray Line Signals from
  Dark Matter Annihilations in the Inner Galactic Halo from 10 Years of
  Observations with H.E.S.S.}},
  \href{https://doi.org/10.1103/PhysRevLett.120.201101}{\emph{Phys. Rev. Lett.}
  {\bfseries 120} (2018) 201101}
  [\href{https://arxiv.org/abs/1805.05741}{{\ttfamily 1805.05741}}].

\end{thebibliography}\endgroup

\end{document}